\documentclass{article}

\usepackage{bioarxiv}
\usepackage[utf8x]{inputenc} 
\usepackage[T1]{fontenc}    
\usepackage{hyperref}       
\usepackage{url}            
\usepackage{booktabs}       
\usepackage{amsfonts}       
\usepackage{nicefrac}       
\usepackage{microtype}      
\usepackage{graphicx}
\usepackage{wrapfig}
\usepackage{amsmath}
\usepackage{multicol}        
\usepackage[bottom]{footmisc}
\usepackage{newtxtext}   
\usepackage{newtxmath}
\usepackage{makecell}
\usepackage{adjustbox}

\usepackage{subfig}
\usepackage{tabularx}
\usepackage{longtable}
\usepackage{multirow}
\usepackage{tabularx}
\usepackage{tablefootnote}
\usepackage{lscape}
\usepackage{setspace} 

\title{Distributed Computing in a Pandemic: A Review of Technologies available for Tackling COVID-19}

\author{
  Jamie J. Alnasir\\
  Department of Computing\\
  Imperial College\\
  London, UK \\
  \texttt{j.alnasir@imperial.ac.uk} \\
}

\begin{document}

\maketitle

\begin{abstract}
The current COVID-19 global pandemic caused by the SARS-CoV-2 betacoronavirus has resulted in over a million deaths and is having a grave socio-economic impact, hence there is an urgency to find solutions to key research challenges. Some important areas of focus are:  vaccine development, designing or repurposing existing pharmacological agents for treatment by identifying druggable targets, predicting and diagnosing the disease, and tracking and reducing the spread. Much of this COVID-19 research depends on distributed computing.

In this article, I review distributed architectures --- various types of clusters, grids and clouds --- that can be leveraged to perform these tasks at scale, at high-throughput, with a high degree of parallelism, and which can also be used to work collaboratively.

High-performance computing (HPC) clusters, which aggregate their compute nodes using high-bandwidth networking and support a high-degree of inter-process communication, are ubiquitous across scientific research --- they will be used to carry out much of this work. Several bigdata processing tasks used in reducing the spread of SARS-CoV-2 require high-throughput approaches, and a variety of tools, which Hadoop and Spark offer, even using commodity hardware.

Extremely large-scale COVID-19 research has also utilised some of the world's fastest supercomputers, such as IBM's SUMMIT --- for ensemble docking high-throughput screening against SARS-CoV-2 targets for drug-repurposing, and high-throughput gene analysis --- and Sentinel, an XPE-Cray based system used to explore natural products. Grid computing has facilitated the formation of the world's first Exascale grid computer. This has accelerated COVID-19 research in molecular dynamics simulations of SARS-CoV-2 spike protein interactions through massively-parallel computation and was performed with over 1 million volunteer computing devices using the Folding@home platform.

Grids and clouds both can also be used for international collaboration by enabling access to important datasets and providing services that allow researchers to focus on research rather than on time-consuming data-management tasks.

\end{abstract}


\section{Introduction}

A novel betacoronavirus named SARS-CoV-2 (Severe Acute Respiratory Syndrome coronavirus 2) is the cause of the clinical disease COVID-19 --- its spread is responsible for the current coronavirus pandemic and the resulting global catastrophe \cite{lake2020we}. The initial outbreak of the disease was first detected in December 2019 in Wuhan (Hubei province, China) manifesting as cases of pneumonia, initially of unknown aetiology. On the 10th of January 2020, Zhang et al. released the initial genome of the virus \cite{zhang2020initial}.  Shortly after, it was identified --- by deep sequencing analysis of lower respiratory tract samples --- as a novel betacoronavirus and provisionally named 2019 novel coronavirus (2019-nCoV) \cite{lu2020outbreak, huang2020clinical}. By the 30th of January 2020, the WHO (World Health Organisation) declared the outbreak a Public Health Emergency of International Concern \cite{who2020statement}, and a global pandemic on the 11th of March \cite{who2020pandemic}. At the time of writing (May 2021) there now are over 160 million reported cases of COVID-19 globally and more than 3,334,000 deaths have occurred as a result of the disease \cite{dong2020interactive}. In addition to the casualties, the pandemic is also having a grave socio-economic impact; it is a global crisis to which researchers will typically apply a variety computational techniques and technologies to several key areas of focus \cite{zhang2020financial, nicola2020socio}. These include, but are not limited to, vaccine development, designing or repurposing existing pharmacological agents for treatment by identifying druggable targets, predicting and diagnosing the disease, e.g. clinical decision support, and tracking and reducing the spread \cite{ferretti2020quantifying, kissler2020projecting, perez2020ongoing}. Many of the tasks involved can leverage a variety of distributed computing approaches which can be applied at scale, at high-throughput, and with a high degree of parallelism --- they often also need to be performed collaboratively.

The classification of SARS-CoV-2 as a betacoronavirus, and the release of its genome earlier on January 2020, has enabled research to focus on specific strategies. It is known from the previous 2003 SARS outbreak that ACE2 (Angiotensin Converting Enzyme) is the main entry point the virus targets to infect its host \cite{li2003angiotensin, kuba2005crucial}. To this end, for drug repurposing or development, COVID-19 research is focused on modelling the interaction between the coronavirus spike protein (S-protein) and ACE2, and in understanding the structure of the S-protein as an epitope for vaccine development. Other important targets are the virus's proteome and the Papain-like and Main proteases -- \textit{PL-pro }and \textit{ML-pro,} respectively \cite{hilgenfeld2014sars}. Given the urgency to reduce mortality, significant efforts are being made to re-purpose medicines that are appropriate and already approved. Whilst a WHO scientific briefing  refers to this practice --- \textit{off-label prescribing} --- in the clinical setting, much of the initial work to predict potential drug candidates will be carried out computationally via \textit{in-silico} screening \cite{kalil2020treating, urlWHOrepurposing}. Furthermore, the scale of the pandemic and the global production of bigdata, particularly whilst vaccines are still being developed, will rely on bigdata analytics to model the spread of the disease, and inform government policy and reduce the death rate.

This review paper explores the variety of distributed and parallel computing technologies which are suitable for the research effort to tackle COVID-19, and where examples exist, work carried out using them. This review will not cover machine-learning or deep-learning methods, which although they employ highly-parallel GPU computing, are not necessarily distributed - whilst they are highly relevant to several COVID-19 research areas, it is separate area in its own right.

The distributed computing architectures that are suitable for application to COVID-19 research exist in several different topologies  --- they can be fundamentally categorised as \textit{clusters}, \textit{grids} and \textit{clouds} \cite{hussain2013survey}, and will be covered in the next sections.

\section{Cluster Computing}\label{lblClusters}

Cluster computing, unlike client-server or n-tier architectures --- where the focus is on delineating resources  --- group together compute nodes to improve performance through concurrency \cite{coulouris2005distributed}. The increasing amount of COVID-19 research to be completed on such systems, coupled with its urgency, will necessitate further performance increases for large-scale projects. They can be achieved by \textit{vertical scaling} --- increasing the number of CPU cores in individual compute nodes of the system --- or \textit{horizontal scaling} --- increasing the number of compute nodes in the system, hence some of the distributed systems employed in the research we review here exhibit both of these characteristics, often at very large-scales.

\subsection{High-performance Computing with MPI}\label{HPC}
High-performance computing (HPC) is a key enabling technology for scientific and industrial research \cite{urlEPSRC-HPC}; HPC systems are ubiquitous across scientific and academic research institutions. Most of the computational research projects  investigating the structure, function and genome of SARS-CoV-2 will be performed on HPC. This will be predominantly in-house, but in some cases will be via access to external HPC resources, e.g. in collaboration between institutions.

By employing high-bandwidth networking interconnects, HPC clusters facilitate a high degree of inter-process communication and extremely large scalability, for instance the Message Passing Interface (MPI) framework. Software implemented using MPI can exploit for many of the computationally complex problems in COVID-19 research, such as ensemble docking and mathematical modeling \cite{hill2000HITreadings}. The use of MPI for distributing complex scientific computation is well-established and many HPC systems are dependent on MPI libraries such OpenMPI and MVAPICH. Consequently, there have been further developments and refinement of these libraries over the last two decades  --- mainly in reducing latency and memory requirements \cite{shipman2006infiniband}. OpenMPI and MPICH have had their most recent releases in 2020. Recently, new MPI implementations are coming to the fore, such as in LinaLC, a docking program employing strategies such as mixed multi-threading schemes to achieve further performance gains at an extremely large scale, that can be applied to COVID-19 research which we will discuss in the next section.

\subsubsection{Ensemble Docking}

A key task in identifying potential pharmacological agents to target SARS-CoV-2 is molecular docking --- \textit{in-silico} simulation of the electrostatic interactions between a ligand and its target --- is used to score ligands according to their affinity to the target \cite{morris2008molecular, meng2011molecular}. The complex computational process is extensively used in drug development and repurposing and is often time-consuming and expensive \cite{moses2005financial, rawlins2004cutting}. The protein and enzyme targets that are docked against are not static, but are constantly moving in ways which are dependent on several factors such as temperature, electrostatic attractions and repulsions with nearby molecules, solvation (interaction with the solvent environment) etc. These factors cause atoms in the molecules, within the constraints of the types of bonds the bind them, to adopt spatial arrangements --- termed conformations --- that correspond to local energy minima on the energy surface. Molecular Dynamics (MD) uses \textit{in-silico} computation to simulate this process, the outcome of which is typically clusters (\textit{"ensembles"}) of the most probable conformations for docking, i.e. ensemble docking \cite{amaro2018ensemble}.

In the past, popular tools such as AutoDock Vina --- widely-used for performing both molecular docking and virtual screening --- were being used primarily on single high-end workstations. Consequently, their parallelism was optimised for multithreading on multicore systems. However, further gains in such tools have been made by developing or re-implementing existing code for the fine-grained parallelism offered by MPI, and at the same time, leveraging the scale at which HPC systems can operate. In previous work, Zhang et al. have further modified the AutoDock Vina source to implement a mixed MPI and multi-threaded parallel version called VinaLC. They have demonstrated this works efficiently at a very large-scale --- 15K CPUs --- with an overhead of only 3.94\%. Using the DUD dataset (Database of Useful Docking Decoys), they performed 17 million flexible compound docking calculations which were completed on 15,408 CPUs within 24 h. with 70\% of the targets in the DUD data set recovered using VinaLC. Projects such as this can be repurposed and applied to identifying potential leads for binding to the SARS-CoV-2 S-protein or the S-protein:Human ACE2 interface, either through the repurposing or the identification of putative ligands \cite{zhang2013message}. Furthermore, given the urgency in finding solutions to the current COVID-19 pandemic --- where high-throughput performance gains and extreme scalability are required --- these features can be achieved by re-implementing tools in similar ways to which VinaLC has been optimised from the AutoDock codebase.

\subsection{Supercomputers and COVID-19}\label{lblSupercomputers}

\subsubsection{Drug Repurposing}

In recent COVID-19 focused research, Smith et al. have utilised IBM's SUMMIT supercomputer --- the world's fastest between November 2018 and June 2020 --- to perform ensemble docking virtual high-throughput screening against both the SARS-CoV-2 S-protein and the S-protein:Human ACE2 interface \cite{smith2020repurposing, urlSUMMITannouncment}.

SUMMIT, launched by ORNL (Oak Ridge National Laboratory) and based at it's Oak Ridge Leadership Computing Facility, comprises 4,608 compute nodes, each with two IBM POWER9 CPUs (containing nine cores each), and six Nvidia Tesla Volta GPUs for a total of 9,216 CPUs and 27,648 GPUs \cite{vazhkudai2018design}.  Nodes each have 600 GB of memory, addressable by all CPUs and GPUs,  with an additional 800 GB of non-volatile RAM that can be used as a burst buffer or as extended memory. SUMMIT implements a heterogeneous computing model --- in each node the two POWER9 CPUs and Nvidia Volta GPUs are connected using Nvidia's high-speed NVLink. The interconnect between the nodes consist of 200 Gb/s Mellanox EDR Infiniband for both storage and inter-process messaging and supports embedded in-network acceleration for MPI and SHMEM/PGAS.

For a source of ligands, they used the SWEETLEAD dataset which is a highly-curated \textit{in-silico} database of 9,127 chemical structures representing approved drugs, chemical isolates from traditional medicinal herbs, and regulated chemicals, including their stereoisomers\cite{novick2013sweetlead}. The work involved three phases of  computation: structural modelling, molecular dynamics simulations (ensemble building), and \textit{in-silico} docking. Since the 3D structure of the SARS-CoV-2 S-protein was not yet available during the initial phase of this research, the first phase (structural modelling) was required and a 3D model was built with SWISSMODEL \cite{schwede2003swiss} using the sequences for the COVID-19 S-protein (NCBI Ref. Seq: YP\_009724390.1) and the crystal structure of SARS-CoV S-protein as a template to generate the model of the SARS-CoV-2 S-protein:ACE2 complex.  In the second phase, molecular dynamics simulations were carried out using GROMACS (compiled on ORNL SUMMIT and run with CHARMM36 force-field \cite{ossyra2019porting, abraham2015gromacs}) to generate an ensemble of highest probability, lowest energy conformations of the complex which were selected via clustering of the conformations. In the final in-silico docking phase, AutoDock Vina was run in parallel using an MPI wrapper.

This work has identified 47 hits for the S-protein:ACE2 interface, with 21 of these having US FDA regulatory approval and 30 hits for the S-protein alone, with 3 of the top hits having regulatory approval.

\subsubsection{High-throughput and Gene Analysis}

Another research project by Garvin et al., that has also been undertaken using SUMMIT, focused on the role of bradykinin and the RAAS (Renin Angiotensin Aldosterone System) in severe, life-threatining COVID-19 symptoms by analysing 40,000 genes using sequencing data from 17,000 bronchoalveolar lavage (BAL) fluid samples \cite{garvin2020mechanistic, urlSUMMITbradykinin}. RAAS regulates blood pressure and fluid volume through the hormones renin, angiotensin and aldosterone. Key enzymes in this system are ACE (Angiotensin Converting Enzyme), and ACE2 which work in antagonistic ways to maintain the levels of bradykinin, a nine-amino acid peptide that regulates the permeability of the veins and arterioles in the vascular system. Bradykinin induces hypotension (lowering of blood pressure) by stimulating the dilation of aerterioles and the constriction of veins, resulting in leakage of fluid into capillary beds. It has been hypothesised that dysregulated of bradykinin signaling is responsible for the respiratory complications seen in COVID-19 --- the \textit{bradykinin storm} \cite{roche2020hypothesized}.

This work involved massive-scale, gene-by-gene RNA-Seq analysis of SARS-CoV2 patient samples with those of the control samples, using a modified transcriptome. The modified transcriptome was created to allow the researchers to quantify the expression of SARS-CoV2 genes and compare them with the expression of human genes. To create the modified transcriptome, reads from the SARS-CoV2 reference genome were appended to transcripts from the latest human transcriptome, thereby allowing the mapping of reads to the SARS-CoV2 genes. The SUMMIT supercomputer enabled the exhaustive gene-wise tests (all the permutations of all the genes) to be performed at a massive scale in order to test for differential expression, with the Benjamini-Hochberg method applied to the resulting p-values to correct for multiple comparisons.

Their analysis appears to confirm dysregulation of RAAS, as they found decreased expression of ACE together with increased expression of ACE2, renin, angeiotensin, key RAAS receptors, and both bradykinin receptors. They also observed increased expression of kininogen and a number of kallikrein enzymes that are kininogen activating --- the activated form, kinins, are polypeptides that are involved in vasodilation, inflammatory regulation, and blood coagulation. As they point out, atypical expression levels for genes encoding these enzymes and hormones are predicted to elevate bradykinin levels in multiple tissues and organ systems, and explain many of the symptoms observed in COVID-19.

\subsubsection{Exploring Natural Products for Treatment}

So far, we have discussed some examples of the use of \textit{in-silico} docking and screening that have utilised HPC to identify existing medicines that could potentially be re-purposed for treating COVID-19. However, another strategy --- one that is used to develop new therapeutics --- explores the chemistry of natural products, i.e. chemical compounds produced by living organisms. To this end, in another research project that also performs in-silico docking and screening using a supercomputer, Sentinel, Baudry et al. focus on natural products \cite{byler2020high}. They point out that, natural products, owing to the long periods of natural selection they are subjected to, perform highly selective functions. Their work, therefore, aims to identify pharmacophores (spatial arrangement of chemical functional groups that interact with a specific receptor or target molecular structure) that can be used to develop rationally designed novel medicines to treat COVID-19. In addition to simulating the interaction with the S-protein RBD, they also included those with the SARS-2 proteome (the sum of the protein products transcribed from its genome), specifically the main protease and the papain-like protease enzymes. These are also important targets as they are highly conserved in viruses and are part of the replication apparatus.

Sentinel, the cluster used for this research, is a Cray XC50, a 48-node, single-cabinet supercomputer featuring a massively parallel multiprocesser architecture and is based in the Microsoft Azure public cloud data centre. It has 1,920 physical Intel Skylake cores operating at 2.4GHz with Hyperthreading (HT) / Simultaneous Multi-Tasking (SMT) enabled, therefore providing 3,840 hyperthreaded CPU cores. Each node has 192 GB RAM and they are connected by an Aries interconnect28 in a Dragonfly topology. A Cray HPE ClusterStor-based parallel file system is used, providing 612 TB of shared storage that is mounted on every node.

The development of medicines from natural products is challenging for several reasons: supply of the plant and marine organisms, seasonal variation in the organism, extinction of organism sources, and natural products often occur as mixtures of structurally related compounds, even after fractionation, only some of which are active. Contamination, stability, solubility of the compounds, culturing source microorganisms, 
and cases where synergistic activities require two constituents to be present to display full activity can also present difficulties \cite{li2009drug}. Baudry et al., therefore, performed their screening using a curated database of 423,706 natural products, COCONUT (COlleCtion of Open NatUral producTs) \cite{urlCOCONUT}. COCONUT has been compiled from 117 existing natural product databases for which citations in literature since 2000 exist.

Using molecular dynamics simulation coordinate files for the structures of the S-protein, main protease and the papain-like protease enzymes --- generated with GROMACS \cite{abraham2015gromacs} and made available by Oak Ridge National Laboratory --- they generated an ensemble using the ten most populated confirmation for each, for \textit{in-silico} docking. As AutoDock was used, its codebase was compiled optimised for Sentinel, with some optimisations for Skylake CPU and memory set in the Makefile compiler options.

They performed pharmacophore analysis of the top 500 unique natural product conformations for each target (S-protein, PL-pro, M-pro). Filtering was applied to the list of putative ligands such that there were no duplicate instances of the same compound, they only occurred in the set for a single target, and were deemed to be drug-like using the MOE (Molecular Operating Environment) descriptor \cite{vilar2008medicinal} from the COCONUT dataset. This resulted in 232, 204, and 164 compounds for the S-protein, PL-pro, M-pro, respectively. Of these, the top 100 natural products were superimposed onto their respective predicted binding locations on their binding proteins and those that correctly bind to the correct region (i.e. active site) were subjected to for pharmacophoric analysis. For the S-protein, two clusters of 24 and 73 compounds were found to bind to either side of a loop that interacts with ACE2. For PL-pro, the papain-like protease, again two clusters of 40 and 60 compounds were found to bind to either side of a beta-sheet. Finally, for ML-pro, the main protease, five clusters of binding compounds were found, one cluster in in close proximity to the proteases catalytic site.

The common pharmacophores partaking in these interactions were assessed from the relevant clusters, resulting in a greater understanding of the structure-activity relationship of compounds likely to be inhibitory to the SARS-CoV-2 S-protein, PL-pro, and ML-pro proteases. As a result, several natural product leads have been suggested which could undergo further testing and development, and the pharmacophore knowledge could be used to refine existing leads and guide rational drug design for medicines to treat COVID-19.

\subsection{Hadoop and Spark}

Apache Hadoop is an open-source software ``ecosystem'' comprising a collection of interrelated, interacting projects, distributed platform and software framework that is typically installed on a Linux compute cluster, notwithstanding that it can be installed on a single standalone machine (usually only for the purposes of study or prototyping). Hadoop is increasingly used for bigdata processing \cite{messerschmitt2005software, joshua2013software}. Bigdata --- which we will discuss in more detail with respect to the COVID-19 pandemic --- is characterised as data possessing large volume, velocity, variety, value and veracity --- known as the v's of bigdata \cite{laney01controlling3v, borgman2015big}. A significant portion of bigdata generated during the COVID-19 pandemic will be semi-structured data from a variety of sources. MapReduce is a formalism for programatically accessing distributed data across Hadoop clusters which store and process data as sets of key-value pairs (i.e. tuples) on which Map and Reduce operations are carried out \cite{fish2015computational}. This makes MapReduce particularly useful for processing this semi-structured data and building workflows.

Apache Spark, often viewed as as the successor to Hadoop is a distributed computing framework in its own right, which can be used standalone or can utilise the Hadoop platform's distributed file system (HDFS), and a resource scheduler — typically Apache YARN. Spark, therefore, can also run MapReduce programs (written in Python, Java, or Scala) \cite{shanahan2015large}. It has been designed to overcome the constraints of Hadoop's acyclic data flow model, through the introduction of a distributed data structure --- the Resilient Distributed Dataset (RDD) --- which facilitates the re-usability of intermediate data between operations, in-memory caching, and execution optimisation (known as lazy evaluation) for significant performance gains over Hadoop \cite{zaharia2012resilient}.

\subsection{COVID-19 Bigdata Analytics}

As briefly mentioned earlier, bigdata refers to data sets that, by virtue of their massive size or complexity, cannot be processed or analysed by traditional data-processing methods, and, therefore, usually require the application of distributed, high-throughput computing. Bigdata analytics --- the collection of computational methods that are applied for gaining valuable insight from bigdata --- employs highly specialised platforms and software frameworks, such as Hadoop or Spark. In a paper that focused on AI for bigdata analytics in infectious diseases, which was written over a year before the current COVID-19 pandemic, Wong et al. point out that, in our current technological age, a variety of sources of epidemiological transmission data exist, such as sentinel reporting systems, disease centres, genome databases, transport systems, social media data, outbreak reports, and vaccinology related data \cite{wong2019artificial}. In the early stages of global vaccine roll out, compounded by the difficulty of scaling national testing efforts, this data is crucial for contact tracing, and for building models to understand and predict the spread of the disease \cite{sun2020early}.

Furthermore, given the current COVID-19 pandemic has rapidly reached a global scale, the amount of data produced and the variety of sources is even greater than before. Such data is, in most cases, semi-structured or unstructured and, therefore, requires pre-processing \cite{agbehadji2020review}. The size and rate in which this data is being produced during this pandemic, particularly in light of the urgency, necessitates bigdata analytics to realise the potential it has to aid in finding solutions to arrest the spread of the disease  by, for example, breaking the chain of transmission (i.e. via track-and-trace systems), and informing government policy \cite{bragazzi2020big}.

The Apache Hadoop ecosystem has a several projects ideally suited to processing COVID-19 big data, and by virtue of them all utilising Hadoop's cluster infrastructure and distributed file system, they gain from the scalability and fault-tolerance inherent in the framework. For example, for pre-processing bigdata --- often referred to as cleaning dirty data --- Pig is a high-level data-flow language that can compile scripts into sequences of MapReduce steps for execution on Hadoop  \cite{olston2008pig}. Apache Spark, owing to its in-memory caching and execution optimisations discussed earlier, offers at least two orders of magnitude faster execution than Hadoop alone and, though centred around MapReduce programming, is less constrained to it. Hive \cite{thusoo2009hive} is a data-warehousing framework which has an SQL type query language, HBase \cite{george2011hbase} a distributed scalable database, and Mahout \cite{lyubimov2016apache} can be used for machine-learning and clustering of data.

An example of how Hadoop can be applied to analytics of COVID-19 big data is shown in recent work by Huang et al who have analysed 583,748,902 geotagged tweets for the purposes of reporting on human mobility --- a causal factor in the spread of the disease \cite{huang2020twitter}. In doing so they have demonstrated that bigdata captured from social media can be used for epidemiological purposes and can do so with less invasion of privacy that such data offers. They do point out, however, that a limitation to this approach is that only a small portion of the total twitter corpus is available via the API. That said, an important outcome of this work is their proposed metric for capturing overall mobility during phases of pandemics --- the MRI (Mobility-Responsiveness) Indicator which can be used as a proxy for human mobility.

Whilst Hadoop and Spark are frequently applied to data analytics, they have also been employed in bioinformatics --- such as in processing Next-generation sequencing data, e.g. SNP genotyping, de novo assembly, read alignment, reviewed in \cite{taylor2010overview} and structural biology, e.g. \textit{in-silico} molecular docking, structural alignment / clustering of protein-ligand complexes, and protein analysis reviewed in \cite{alnasir2020application}.

\section{Grid Computing}\label{lblGrids}

Grids provide a medium for pooling resources and are constructed from a heterogeneous collection of geographically dispersed compute nodes connected in a mesh across the internet or corporate networks. With no centralised point of control, grids broker resources by using standard, open, discoverable protocols and interfaces to facilitate dynamic resource-sharing with interested parties \cite{foster2008cloud}. Particularly applicable to COVID-19 research are the extremely large scalability grid computing offers and the infrastructure for international collaboration they facilitate, which we will discuss in the following sections.

\subsection{Large-scale Parallel-processing Using Grids}

The grid architecture allows for massive parallel computing capacity by the \textit{horizontal} scaling of heterogeneous compute nodes, and the exploitation of underutilised resources through methods such as idle CPU-cycle scavenging \cite{bhavsar2009scavenging}. Distributed, parallel processing using grids is ideally suited for batch tasks that can be executed remotely without any significant overhead. 

An interesting paradigm of grid computing, that has now been applied to Molecular Dynamics research for COVID-19, leverages this scalability, particularly for applications in scientific computing, is known as volunteering distributed computing having evolved during the growth of the internet from the 2000s onwards. This involves allocating work to volunteering users on the internet (commonly referred to as the @home projects) with tasks typically executed while the user's machine is idle \cite{krieger2002models}.

\subsection{The World's First Exascale Computer Assembled Using Grid Volunteer Computing}\label{lblExaScaleGridComputer}

A recent project, that focused on simulating the conformations adopted by the SARS-CoV-2 S-protein, culminated in the creation of the first Exascale grid computer. This was achieved by enabling over a million citizen scientists to volunteer their computers to the Folding@home grid computing platform, which was first founded in 2000 to understand protein dynamics in function and dysfunction \cite{zimmerman2020citizen, beberg2009folding}. The accomplishment of surmounting the Exascale barrier by this work is based on a conservative estimate that the peak performance of 1.01 exaFLOPS on the Folding@home platform was achieved at a point when ~280,000 GPUs and 4.8 million CPU cores were performing simulations. The estimate counts the number of GPUs and CPUs that participated during a three-day window, and makes the conservative assumption about the computational performance of each device. Namely, that each GPU/CPU participating has worse performance than a card released before 2015.

In addition to understanding how the structure of the SARS-CoV-2 S-protein dictates its function, simulating the ensemble of conformations that it adopts allows characterisation of its interactions. These interactions with the ACE2 target, the host system antibodies, as well as glycans on the virus surface, are key to understanding the behaviour of the virus. However, as pointed out in this work, datasets generated by MD simulations typically consist of only a few microseconds of simulation --- at most millisecond timescales --- for a single protein. An unprecedented scale of resources are therefore required to perform MD simulations for all of the SARS-CoV-2 proteins. The Folding@home grid platform has enabled this, generating a run of 0.1 s of simulation data that illuminates the movement and conformations adopted by these proteins over a biologically relevant time-scale.

\subsection{International Collaboration Through Grids}\label{lblCollabGrids}

Grids are an ideal infrastructure for hosting large-scale international collaboration. This was demonstrated by the Globus Toolkit produced by the Global Alliance, which became a de facto standard software for grids deployed in scientific and industrial applications. It was designed to facilitate global sharing of computational resources, databases and software tools securely across corporate and institutions \cite{ferreira2003introduction}. However, development of the toolkit ended in 2018 due to a lack of funding and the service remains under a freemium mode. Globus's work in enabling worldwide collaboration continue through their current platform which now employs cloud computing to provide services --- this is discussed further in section \ref{lblClouds}.

Some notable large-scale grids participating in COVID-19 research are: the World Community Grid launched by IBM \cite{urlGridWCG}, the WLCG (Worldwide LHC Computing Grid) at CERN \cite{sciaba2010computing}, Berkeley Open Infrastructure for Network Computing (BOINC) \cite{anderson2004boinc}, the European Grid Infrastructure (EGI) \cite{gagliardi2004egee}, the Open Science Grid (OSG) \cite{pordes2007open} and previously Globus. Interestingly, grids can be constructed from other grids --- for example, BOINC is part of IBM WCG, and CERN's WLCG is based on two main grids, the EGI and OSG, which is based in the US.

\subsection{COVID-19 Research on Genomics England Grid}\label{lblGE}

Genomics England, founded in 2014 by the UK government and owned by the Department of Health \& Social Care, has been tasked with delivering the 100,000 genomes project which aims to study the genomes of patients with cancer or rare diseases \cite{siva2015uk, urlGovGenomesBBC}.  It was conceived at a time when several government and research institutions worldwide announced large-scale sequencing projects --- akin to an \textit{arms race} of sequencing for patient-centric precision medicine research. In establishing the project, the UK government and Illumina decided to secure sequencing services for the project from Illumina \cite{marx2015dna}. Sequencing of the 100,000 genomes has resulted in 21 PB of data and involved 70,000 UK patients and family members, 13 genomic medicines centres across 85 recruiting NHS trusts, 1,500 NHS staff, and 2,500 researchers and trainees globally \cite{urlGovGenomicsEnglandNumbers}.

In 2018, after sequencing of the 100,000 genomes was completed, the UK government announced the significant expansion of the project --- to sequence up to five million genomes over five years \cite{urlGovGenomicsEngland5million}. At the time, the Network Attached Storage (NAS) held 21 PB of data and had reached its node-scaling limit and so a solution that could scale to hundreds of Petabytes was needed --- after consultation with Nephos Technologies, a more scalable storage system comprising a high-performance parallel file system from WekaIO, Mellanox® high-speed networking, and Quantum ActiveScale object storage was implemented \cite{urlGovGenomicsEnglandQuantum}. Genomics England's Helix cluster, recently commissioned in 2020, has 60 compute nodes each with 36 cores (providing 2,160 cores) and approximately 768 GB RAM. It has a dedicated GPU node with 2x nVidia Tesla V100 GPUs installed \cite{urlGovGenomicsEnglandHelix}.

GenOMICC (Genetics Of Mortality In Critical Care) is a collaborative project, first established in 2016, to understand and treat critical illness such as sepsis and emerging infections (e.g. SARS/MERS/Flu) is now also focusing on the COVID-19 pandemic. The collaboration involves Genomics England, ISARIC (The International Severe Acute Respiratory and Emerging Infection Consortium), InFACT (The International Federation of Acute Care Triallists), Asia-Pacific Extra-Corporeal Life Support Organisation (AP ELSO) and the Intensive Care Society. The aim is to recruit 15,000 participants for genome sequencing, who have experienced only mild symptoms, i.e. who have  tested positive for COVID-19, but have not been hospitalised. The rationale is that in addition to co-morbidities, there are genetic factors that determine whether a patient will suffer mild or severe, potentially life-threatenting illness --- this would also explain why some young people, who are fit and healthy have suffered severely and others who are old and frail did not. Furthermore, since many people who have suffered severe illness from COVID-19 were elderly or from ethnic minorities, the aim is to recruit participants that are from these backgrounds who suffered from mild symptoms of COVID-19. To this end, the project will carry out GWAS (Genome Wide Association Studies) to identify associations between genetic regions (loci) and increased susceptibility to COVID-19 \cite{pairo2020genetic}.

\subsection{Other COVID-19 Research on Grids}

During the previous 2002-4 SARS-CoV-1 outbreak, DisoveryNet --- a pilot designed and developed at Imperial College and funded by the UK e-Science Programme --- enabled a collaboration between its team and researchers from SCBIT (Shanghai Centre for Bioinformation Technology) to analyse the evolution of virus strains from individuals of different countries \cite{au2004grid}. This was made possible through its provision of computational workflow services, such as an XML based workflow language and the ability to couple workflow process to datasources, as part of an e-Science platform to facilitate the extraction of knowledge from data (KDD)  \cite{rowe2003discovery}. It is coincidental that this grid technology in its infancy, and in its pilot phase, was used in a prior pandemic, especially since many of the services will be employed during the current one, in particular the support for computational workflows and the use of large datasets made available through grids and clouds.

In work that utilises various grid resources, including EGI and OSG, together with the European Open Science Cloud (EOSC) \cite{ayris2016realising}, Hassan et al. have performed an \textit{in-silico} docking comparison between human COVID-19 patient antibody (B38) and RTA-PAP fusion protein (ricin a chain-pokeweed antiviral protein) against targets (S-protein RBD, Spike trimer, and membrane-protein) in SARS-CoV-2 \cite{hassan2020novel}. RTA-PAP, plant-derived N-glycosidase ribosomal-inactivating proteins (RIPs), is a fusion of ricin a chain --- isolated from Ricinus communis --- and pokeweed antiviral protein --- isolated from Phytolacca Americana, which the same researchers had demonstrated to be anti-infective against Hepatitis B in prior work \cite{hassan2018expression}. They also utilised a grid based service called WeNMR, which provides computational workflows for NMR (Nucleic Magnetic Resonance)/SAX (Small-angle X-ray scattering)  via easy-to-use web interfaces \cite{wassenaar2012wenmr}, and the CoDockPP protein-protein software to perform the docking \cite{kong2019codockpp}. They found favourable binding affinities (low binding energies) for the putative fusion protein RTA-PAP binding with both the SARS-CoV-2 S-protein trimer and membrane protein, which can be further explored for development as antivirals for use against COVID-19.

\section{Cloud Computing}\label{lblClouds}

A consequence of the data-driven, integrative nature of bioinformatics and computational biology \cite{dudley2010translational}, as well as advancements in high-throughput next-generation sequencing \cite{dai2012bioinformatics}, is that cloud-services such as for instance Amazon AWS, Microsoft Azure, and Google Cloud, are increasingly being used in research \cite{schatz2010cloud, shanahan2014bioinformatics}. These areas of research underpin the COVID-19 research effort and hence the use of cloud services will no doubt contribute significantly to the challenges faced.

Clouds provide pay-as-you-go access to computing resources via the internet through a service provider and with minimal human interaction between the user and service provider. Resources are accessed on-demand, generically as a service, without regard for physical location or specific low-level hardware and in some cases without software configuration \cite{smith2005architecture}. This has been made possible by the developments in virtualisation technologies such as Xen, and Windows Azure Hypervisor (WAH) \cite{younge2011analysis, barham2003xen}. Services are purchased on-demand in a metered fashion, often to augment local resources and aid in completion of large or time-critical computing tasks. This offers small research labs access to infrastructure that they would not be able to afford to invest in for use on-premises, as well as services that would be time consuming and costly to develop \cite{navale2018cloud}. Furthermore, there is variety in the types of resources provided in the form of different service models offered by cloud providers, such as Software as a Service (SaaS). Platform as a Service (PaaS) and Infrastructure as a Service (IaaS) \cite{mell2011nist}.

To a great extent, scientific and bioinformatics research projects utilise cloud services through IaaS (Infrastructure as a Service) and PaaS (Platform as a Service). In the IaaS approach, processing, storage and networking resources are acquired and leased from the service provider and configured by the end user to be utilised through the use of virtual disk images. These virtual disks are provided in proprietary formats, for instance the AMI (Amazon Machine Image) on AWS or VHD (Virtual Hard Disk) on Azure, serve as a bit-for-bit copy of the state of a particular VM \cite{shanahan2014bioinformatics}. They are typically provisioned by the service provider with an installation of commonly used Operating Systems configured to run on the cloud service's Infrastructure, and service providers usually offer a selection of such images. This allows the end user to then install and precisely configure their own or third party software, save the state of the virtual machine, and deploy the images elsewhere.

In contrast, in the PaaS approach, the end user is not tasked with low level configuration of software and libraries which are instead provided to the user readily configured to enable rapid development and deployment to the cloud. For example AWS provides a PaaS for MapReduce called Elastic MapReduce which it describes as a ``Managed framework for Hadoop that makes it easy, fast, and cost-effective to process vast amounts of data across dynamically scalable Amazon EC2 instances'' \cite{urlAWSEMR}. In fact MapReduce is offered as a PaaS by all of the major cloud-service providers (Amazon AWS, Google Cloud and Microsoft Azure) \cite{gunarathne2010mapreduce}.

Cloud computing is market-driven and has emerged thanks to improvements in capabilities the internet which have enabled the transition of computational research from mainstay workstations and HPC clusters into the cloud. Clouds offer readily provisionable resources which, unlike grids ---- where investment can be lost when scaled down --- projects utilising cloud infrastructure do not suffer the same penalty. However, there is no up-front cost to establishing infrastructure in the case of clouds. One potential drawback is that whilst the ingress of data into clouds is often free, there is invariably a high cost associated with data egress which, depending on the size of the computational results, can make it more costly than other infrastructures in terms of extricating computational results \cite{navale2018cloud}. These are salient factors with respect to the likely short-term duration of some of the pandemic research tasks that are being carried out.

\subsection{International Collaboration through Clouds}\label{lblCollabClouds}

As discussed in earlier (section \ref{lblCollabGrids}), the Globus service has evolved from the Globus Alliance grid consortium's work on the standardisation and provision of grid services. Currently, Globus is a cloud-enabled platform that facilitates collaborative research through the provision of services which focus primarily on data management. Globus is used extensively by the global research community, and at the time of writing, there are over 120,000 users across more than 1,500 institutions registered and connected by more than 30,000 global \textit{endpoints}. Endpoints are logical file transfer locations (source or destination) that are registered with the service and represent a resource (e.g. a server, cluster, storage system, laptop, etc.) between which files can be securely transferred by authorised users. Globus has enabled the transfer of more than 800 PB of data to-date --- presently more than 500 TB are transferred daily. Some of the services it provides are given in Table \ref{tab:tblGlobusServices} below.

\begin{table}[ht!]
    \centering
    \scalebox{0.92}{

      \begin{tabular}{p{3.5cm}p{7.5cm}}
        \textbf{Feature}  & \textbf{Description} \\ \hline
        \textbf{Identity management}  & {Authentication and authorization interactions are brokered between end-users, identity providers, resource servers (services), and clients} \\
        \addlinespace[.1cm]
        \textbf{File transfers} & {Can be performed securely, either by request or automated via script} \\
        \addlinespace[.1cm]
        \textbf{File sharing} & {Allows sharing between users, groups, and setting access permissions} \\
        \addlinespace[.1cm]
        \textbf{Workflow automation} & {Automate workflow steps into pipelines} \\
        \addlinespace[.1cm]
        \textbf{Dataset assembly} & {Researchers can develop and deposit datasets, and describe their attributes using domain-specific metadata.} \\
        \addlinespace[.1cm]
        \textbf{Publication repository} & {Curators review, approve and publish data}  \\
        \addlinespace[.1cm]
        \textbf{Collaboration} & {Collaborators can access shared files via Globus login  --- no local account is required --- and then download}  \\
        \addlinespace[.1cm]
        \textbf{Dataset discovery} & {Peers and collaborators can search and discover datasets}  \\ \hline
       
      \end{tabular}}
\caption[]{\label{tab:tblGlobusServices}Some important services Globus provides}
\end{table}

Another cloud project to enable research collaboration, currently still in development, is the European Open Science Cloud (EOSC), which was proposed in 2016 by the European Commission with a vision to enable Open Science \cite{ayris2016realising}. It aims to provide seamless cloud-services for storage, data management and analysis and facilitate re-use of research data by federating existing scientific infrastructures dispersed across EU member states. After an extensive consultation period with scientific and institutional stakeholders, the outcome is a road-map of project milestones, published in 2018, these are: i) Architecture, ii) Data, iii) Services, iv) Access and Interface, and v) Rules and Governance --- and are anticipated to be completed by 2021.

\section{Distributed Computing Resources provided freely to COVID-19 Researchers}

In order to facilitate and accelerate COVID-19 research, a number of organisations are offering distributed computational resources freely to researchers \cite{urlFreeCOVID19resources}. For instance, a number of research institutions that extensively use HPC --- many of which host the world's most powerful supercomputers --- have joined together to form the COVID-19 High Performance Computing Consortium. Cloud providers such as Amazon AWS, Google, Microsoft and Rescale are also making their platforms available, generally through the use of computational credits, and are largely being offered to researchers working on COVID-19 diagnostic testing and vaccine research. Table \ref{tab:tblFreeServices} lists some of the computational resources on offer, and the specific eligibility requirements for accessing them.

\begin{table}
    \centering
    \scalebox{0.82}{
    \begin{tabular}{p{5cm}p{5cm}p{7cm}} \toprule
\textbf{Provider / Initiative} & \textbf{Offering} & \textbf{Eligibility} \\
\midrule

\cite{urlFreeCOVID19HPCConsortium} COVID-19 High Performance Computing Consortium & {Access to global supercomputers at the institutions taking part in the consortium, such as:

\begin{itemize}
    \item Oak Ridge Summit
    \item Argonne Theta
    \item Lawrence Berkeley National Laboratory Cori
    \item and many more
\end{itemize}
Also, other resources contributed by members, such as:
\begin{itemize}
    \item IBM, HP, Dell, Intel, nVidia
    \item Amazon, Google
    \item National infrastructures (UK, Sweden, Japan, Korea, etc)
    \item and many others
\end{itemize}
} &

{Requests need to demonstrate:
\begin{itemize}
\item Potential near-term benefits for COVID-19 response
\item Feasibility of the technical approach
\item Need for HPC
\item HPC knowledge and experience of the proposing team
\item Estimated computing resource requirements 
\end{itemize}
} \\

\cite{urlFreeCOVID19Rescale} Tech against COVID: Rescale partnership with Google Cloud and Microsoft Azure & {HPC resources through the Rescale platform} & {Any researcher, engineer, or scientist can apply who is targeting their work to combat COVID-19 in developing test kits and vaccines.} \\

\addlinespace[.3cm]

\cite{urlFreeCOVID19AmazonDDI} AWS Diagnostic Development Initiative\tablefootnote{Although the size of the infrastructure in this project is small, the dataset represents a large-scale study.} & {AWS in-kind credits and technical support.} & { Accredited research institutions or private entities:
\begin{itemize}
    \item a using AWS to support research-oriented workloads for the development of point-of-care diagnostics
    \item other COVID-19 infectious disease diagnostic projects considered
\end{itemize} } \\

\cite{urlFreeCOVID19Lifebit} Lifebit & {Premium license for Lifebit CloudOS} & {Exact eligibility criteria not published, but is advertised for researchers developing diagnostics, treatments, and vaccines for COVID-19. Contact lifebit with details of project.} \\

\addlinespace[.3cm]

\bottomrule
\end{tabular}}
    \caption[]{\label{tab:tblFreeServices}Free HPC and Cloud-computing resources for COVID-19 researchers}
    
\end{table}

\clearpage
\begin{table}
    \centering
    \scalebox{0.72}{
    \begin{tabular}{p{2.5cm}p{2.7cm}p{3.35cm}p{4.3cm}p{2.7cm}p{5.5cm}} \toprule
\textbf{Ref. / Name} & \textbf{Platform} & \textbf{Scale} & \textbf{Research task} & \textbf{Tools} & \textbf{Outcome} \\
\midrule
\cite{smith2020repurposing} Smith et al. & \makecell[l]{IBM Summit\\supercomputer} & \makecell[l]{up to 4,608  nodes,\\9,216 CPUs,\\27,648 GPUs} & \textit{in-silico} ensemble docking \& screening of existing medicines for repurposing & GROMACS, CHARMM32, AutoDock Vina & Identified 47 hits for the S-protein:ACE2 interface, with 21 of these having US FDA regulatory approval. 30 hits for the S-protein alone, with 3 of the top hits having regulatory approval. \\

\cite{garvin2020mechanistic} Garvin et al. & \makecell[l]{IBM Summit\\supercomputer} & \makecell[l]{up to 4,608  nodes,\\9,216 CPUs,\\27,648 GPUs} & large-scale gene analysis & AutoDock & Observed atypical expression levels for genes in RAAS pointing to bradykinin dysregulation and storm hypothesis. \\

\cite{byler2020high} Baudry et al. & \makecell[l]{Cray Sentinel\\supercomputer} & \makecell[l]{up to 48 nodes,\\ 1,920 physical cores\\3,840 HT/SMT cores} & \textit{in-silico} docking &  AutoDock & Pharmacophore analysis of natural product compounds likely to be inhibitory to the SARS-CoV-2 S-protein, PL-pro, and ML-pro proteases. \\
\addlinespace[.3cm]

\cite{zimmerman2020citizen} Zimmerman et al. & \makecell[l]{Folding@home grid} & \makecell[l]{4.8 million CPU cores\\$\sim$280,000 GPUs} & MD simulations & GROMACS, CHARMM36, AMBER03 &
Generated an unprecedented 0.1 s of MD simulation data. \\

\addlinespace[.3cm]
\cite{hassan2020novel} Hassan et al. & \makecell[l]{EGI, OSD grids\\\& EOSC cloud} & \textit{unspecified} & \textit{in-silico} docking &
weNMR, CoDockPP &
Demonstrated high in-silico binding affinities of fusion protein RTA-PAP putative ligand with both the SARS-CoV-2 S-protein trimer and membrane protein. \\

\addlinespace[.3cm]
\cite{pairo2020genetic} Pairo et al. & \makecell[l]{Genomics England\\ grid, Helix cluster} & \makecell[l]{up to 60 nodes\\(2,160 cores),\\ 2x V100 GPUs} & GWAS &
Not yet specified &
Recruitment of 15,000 participants is ongoing. \\

\addlinespace[.3cm]
\cite{huang2020twitter} Huang et al. & \makecell[l]{On-premises\\ Hadoop cluster} &
\makecell[l]{13 Hadoop nodes \tablefootnote{Although the size of the infrastructure in this project is small, the dataset represents a large-scale study.}} &
Twitter analytics &
Hive, Impala &
Analysis of over 580 million global geo-tagged tweets demonstrated that twitter data is amenable to quantitatively assess user mobility for epidemiological study, particularly in response to periods of the pandemic and government announcements on mitigating measures. Metric proposed: MRI (Mobility-based Response Index) to act as proxy for human movement. \\

\addlinespace[.3cm]

\bottomrule
\end{tabular}}
    \caption[]{\label{tab:tblComparison}Comparison of COVID-19 research exploiting large-scale distributed computing}
    
\end{table}

\section{Conclusions}

There are a variety of distributed architectures that can be employed to perform efficient, large-scale, and highly-parallel computation requisite for several important areas of COVID-19 research. Some of the large-scale COVID-19 research projects we have discussed that utilise these technologies are summarised in Table \ref{tab:tblComparison} --- these have focused on \textit{in-silico} docking, MD simulation and gene-analysis.

High-performance computing (HPC) clusters are ubiquitous across scientific research institute and aggregate their compute nodes using high-bandwidth networking interconnects. Employing communications protocols, such as Message Passing Interface (MPI), they enable software to achieve a high degree of inter-process communication. Hadoop and Spark facilitate high-throughput processing suited for the bigdata tasks in COVID-19 research. Even when Hadoop/Spark clusters are built using commodity hardware, their ecosystem of related software projects can make use of the fault-tolerant, scalable Hadoop framework i.e. HDFS distributed file system --- features that are usually found in more expensive HPC systems. Although not widely adopted, nor a common use, Hadoop and Spark have also been employed for applications in bioinformatics (e.g. processing sequencing data) and structural biology (e.g. performing docking, clustering of protein-ligand conformations).

\textbf{Key points}
\begin{itemize}
\item HPC is commonly used in research institutions. However, access to the world's supercomputers allows for the largest scale projects to be completed quicker, which is particularly important given the time urgency of COVID-19 research.

\item Bigdata generated during the pandemic - which can be used for epidemiological modeling and critical track and trace systems - can be processed using platforms such as Spark and Hadoop.

\item Grid computing platforms offer unprecedented computing power through volunteer computing, enabling large-scale analysis during the pandemic that hitherto has not been achieved at this scale.

\item Both grids and clouds can also be used for international research collaboration by providing services, frameworks and APIs, but differ in their geographical distribution and funding models.
\end{itemize}

COVID-19 research has utilised some of the world's fastest supercomputers, such as IBM's SUMMIT --- to perform ensemble docking virtual high-throughput screening against SARS-CoV-2 targets for drug-repurposing, and high-throughput gene analysis --- and Sentinel, an XPE-Cray based system used to explore natural products. During the present COVID-19 pandemic, researchers working on important COVID-19 problems, who have relevant experience, now have expedited and unprecedented access to supercomputers and other powerful resources through the COVID-19 High Performance Computing Consortium. Grid computing has also come to the fore during the pandemic by enabling the formation of an Exascale grid computer allowing massively-parallel computation to be performed through volunteer computing using the Folding@home platform.

Grids and clouds provide services such as Globus provide a variety of services, for example, reliable file transfer, workflow automation, identity management, publication repositories, and dataset discovery, thereby allowing researchers to focus on research rather than on time-consuming data-management tasks. Furthermore, cloud providers such as AWS, Google, Microsoft and Rescale are offering free credits for COVID-19 researchers.

In the near future, we will be able to assess the ways in which distributed computing technologies have been deployed to solve important problems during the COVID-19 pandemic and we will no doubt learn important lessons that are applicable to a variety of scenarios.

\section{Acknowledgements}

The author wishes to thank Eszter Ábrahám for proofreading the manuscript.

\section{Conflicts of Interest Statement}

The author declares no conflicts of interest.

\section{Authors' information}

Jamie J. Alnasir is currently a post-doctoral research associate at the SCALE Lab, Department of Computing, Imperial College London. Having gained his Ph.D. from the University of London, he worked in the Scientific Computing department at the Institute of Cancer Research in London providing consulting on HPC and Research Software Engineering before moving to Imperial College. His research interests are distributed computing, high-performance computing, DNA-storage, computational biology, next-generation sequencing and bioinformatics. He is also a Genomics England Clinical Interpretation Partnership (GeCIP) member.

\bibliographystyle{unsrt}  
\bibliography{main.bbl} 

\end{document}